# A Self-Organizing Multi-Agent System for Distributed Voltage Regulation

Badr Al Faiya, *Student Member, IEEE*, Dimitrios Athanasiadis, *Member, IEEE*, Minjiang Chen, Stephen McArthur, *Fellow, IEEE*, Ivana Kockar, *Senior Member, IEEE*, Haowei Lu, and Francisco de León, *Fellow, IEEE*

*Abstract*—This paper presents a distributed voltage regulation method based on multi-agent system control and network self-organization for a large distribution network. The network autonomously organizes itself into small subnetworks through the epsilon decomposition of the sensitivity matrix, and agents group themselves into these subnetworks with the communication links being autonomously determined. Each subnetwork controls its voltage by locating the closest local distributed generation and optimizing their outputs. This simplifies and reduces the size of the optimization problem and the interaction requirements. This approach also facilitates adaptive grouping of the network by self-reorganizing to maintain a stable state in response to time-varying network requirements and changes. The effectiveness of the proposed approach is validated through simulations on a model of a real heavily-meshed secondary distribution network. Simulation results and comparisons with other methods demonstrate the ability of the subnetworks to autonomously and independently regulate the voltage and to adapt to unpredictable network conditions over time, thereby enabling autonomous and flexible distribution networks.

*Index Terms*—Distributed generation, distributed voltage regulation, epsilon decomposition, multi-agent systems, self-organization.

## I. INTRODUCTION

THE increasing penetration of distributed and renewable generation sources in smart grids presents distribution networks with various technical challenges such as voltage control, power quality, and grid losses. Additionally, local elements of a power system, including distributed generation (DG), loads, and the network itself, have varying degrees of influence and coupling to each other. These challenges can be solved while still maintaining system-level coordination through the use of distributed and decentralized control methods based on local information [1]–[3]. In this paper, we present a novel distributed voltage regulation technique able to autonomously divide a system into sub-systems while dealing with time-varying conditions.

The voltage regulation on a distribution networks is customarily implemented by regulating devices, such as on-load tap changer (OLTC) substation transformers, capacitor banks, and

Badr Al Faiya, Dimitrios Athanasiadis, Minjiang Chen, Stephen McArthur and Ivana Kockar are with the Electronic & Electrical Engineering Department, University of Strathclyde, Glasgow, G1 1XW, UK (e-mails: {badr.alfaiya, dimitrios.athanasiadis, minjiang.chen, s.mcarthur, ivana.kockar}@strath.ac.uk).

Haowei Lu and Francisco de León are with the Department of Electrical and Computer Engineering, New York University, Brooklyn, NY 11201 USA (e-mails: {haowei.lu, fdeleon}@nyu.edu).

voltage regulators in the feeders. For a distribution network with multiple DGs, however, the settings of these traditional regulating devices are not the same as for a network without DGs [4]. Additionally, these regulating devices do not react fast enough during emergency conditions [4]. A number of publications have proposed distributed voltage control schemes based on DG [4]–[9]. In [5], a distributed control technique for mitigating overvoltage events by controlling the active and reactive power outputs of inverters in feeders was proposed that requires coordination between feeders. The authors of [6] presented a voltage control scheme in which generation curtailment of DG is introduced if the reactive power exceeds certain limits. Additional publications have tackled distributed voltage control schemes either by integrating sensitivity methods [7], [8] or multi-agent systems (MAS) [4], [9]. In [4], an MAS-based voltage support scheme was presented using DGs in a distribution feeder. An agent-based control model was presented in [9] in which the output power of the DGs is adjusted to reach the balance between the supply and demand while providing stability of the voltage and frequency. All the above distributed techniques are able to tackle network challenges; however, none of them is able to provide distributed and autonomous voltage regulation with dynamic grouping of DGs by considering the unanticipated conditions of DGs and the highly dynamic behavior of the emerging smart grids.

The growing number of controllable devices (e.g., DG) and the corresponding control variables, as well as the increasing volume of data and information in future smart distribution networks, are leading to unprecedented complexity in the required control paradigm [1], [2], [10]. Additionally, time-varying network conditions and the availability of DGs over time cannot be easily predicted during design and, thus, flexible techniques that can deal with such evolving control requirements are necessary. The concept of self-organization has attracted interest in the context of distributed and complex systems to address their uncertainties and dynamic requirements [11]. In addition to its well-known benefit of adaptability, self-organization also has key features such as decentralized and dynamic properties [11].

To address these challenges, this paper presents a control scheme based on a self-organizing MAS for distributed voltage regulation using appropriate DGs. The contributions of this paper are as follows. First, this approach enables a decomposition technique in which a large distribution network autonomously self-subdivides into smaller subnetworks, thereby reducing the size of the problem while enabling voltage control in a distributed and cooperative manner. Each subnetwork regulates its voltage autonomously and independently using appropriate DGs in the same subnetwork. Second, the self-organization is enabled through a mechanism that adapts network subdivisions to reflect varying network conditions. The desired control mechanism is resilient to network anomalies and uses local interactions to adjust the structure of the MAS without stopping the system. Third, the effectiveness of the algorithm is tested on a model of a real heavily-meshed distribution network with DGs and compared with other schemes demonstrating its autonomy and robustness under time-varying network conditions.

The paper is organized as follows. Section II reviews the decomposition and self-organization approaches, while Section III describes the self-organizing MAS framework for distributed voltage regulation with details of the system functions and behaviors. The simulation platform and the deployment of the algorithm in the test system are defined in Section IV. To demonstrate the system for voltage violation and regulation scenarios, comparative performance evaluations with other methods are presented in Section V. Section VI provides our conclusions.

## II. DECOMPOSITION AND SELF-ORGANIZATION

An optimal voltage regulation approach was presented in [8] based on the application of the epsilon decomposition method to the sensitivity matrix (inverse of the Jacobian matrix of the Newton-Raphson power flow problem) to group the DG into matrices based on their influence on the voltage. The authors carried out planning studies in advance to test anticipated worst-case scenarios and to determine fixed partitioning to be used in the future. However, this decomposition and voltage regulation method can be realized by means of a distributed control scheme. In this paper, we implement the desired decomposition method in an MAS architecture and extend this method by introducing a dynamic self-partitioning technique to the network to resolve voltage issues.

Recently, some researchers have attempted to explore self-organization for power networks. A power flow control scheme for DGs in feeders was developed in [12] by applying cooperative control theory to coordinate feeders. A self-organizing communication architecture was proposed in [13] using MAS to mitigate cyber-threats on control schemes in smart grids. Moreover, energy market problems, such as an economic dispatch approach based on hierarchical particle swarm optimization [14], adaptive short-term load forecasting using a self-organized map (SOM) [15], and price and performance management for microgrids using an MAS [16], have been widely investigated. Consensus-based protocols have been used in recent studies to implement self-organizing properties in power networks [17]-[21]. Although consensus approaches have promising properties, they require iterative global communication and updates to reach consensus and synchronized results [22], and are mostly applied in small-scale radial networks. In previous studies, the problems of how to control the voltage using appropriate DGs in large networks and how to make groups of DGs autonomous and dynamic have not been investigated.

Through the abovementioned methods, researchers have demonstrated that self-organizing systems exhibit flexible organization behavior to permit the realization of a desired objective while adapting to the conditions of the immediate environment. These techniques can intelligently enable distributed control with autonomous agents, which can facilitate the exploitation of the inherent flexibility of DGs in network operations. In contrast to the methods discussed above, this paper presents a prototype self-organizing MAS control system for distributed voltage regulation in large networks through local interactions of agents in a cooperative way. The proposed algorithm addresses the challenge of enabling distributed control and the self-organization of multiple subnetworks while dealing with uncertainties in the network.

## III. DEPLOYING EPSILON-DECOMPOSITION-BASED CONTROL: SELF-ORGANIZATION THROUGH AN MAS

### A. Epsilon-Decomposition-Based Voltage Regulation

Epsilon ($\varepsilon$) decomposition is an algorithm that breaks up a matrix into diagonal submatrices [23]. The concept is based on the premise that in a given Jacobian matrix $A = (a_{ij})$ and with a threshold $0 < \varepsilon < 1$, all entries with values less than $\varepsilon$ are set to zero. The "new" matrix, which is a block diagonal decomposed matrix, contains all variables that are strongly coupled in the same block, while weak couplings are discarded. The number of discarded weak couplings depends on the $\varepsilon$ value used for decomposition, e.g., for higher values of $\varepsilon$, there are fewer couplings that need to be considered, resulting in smaller groups in terms of the number of variables. In [8], an epsilon decomposition algorithm was applied to the sensitivity matrix to achieve the optimal voltage regulation.

The epsilon decomposition method can be applied to divide a large power system into subsystems. The sensitivity matrix can be derived from the Jacobian matrix of the Newton-Raphson power flow problem as

$$\begin{pmatrix} \Delta\theta \\ \Delta V \end{pmatrix} = \begin{pmatrix} A_{\theta P} & A_{\theta Q} \\ A_{VP} & A_{VQ} \end{pmatrix} \begin{pmatrix} \Delta P \\ \Delta Q \end{pmatrix} \qquad (1)$$

where the sensitivity matrix $A$ is

$$A = \begin{pmatrix} A_{\theta P} & A_{\theta Q} \\ A_{VP} & A_{VQ} \end{pmatrix}. \qquad (2)$$

The sensitivity matrix $A$ describes the linear relationships between the changes in the active and reactive power levels of a DG and the voltage variations. We can calculate adjustments to the DG active power, $x_P = P_r - P_0$, or reactive power, $x_Q = Q_r - Q_0$, to control the voltage at a particular node from an initial voltage $V_0$ to a reference voltage $V_r$ as

$$V_r = V_0 + A_{VP} \cdot x_P + A_{VQ} \cdot x_Q \qquad (3)$$

where $P_0$ and $P_r$ are the DG active power outputs before and after voltage regulation, respectively, and $Q_0$ and $Q_r$ are the reactive power outputs before and after voltage regulation, respectively.

The purpose of epsilon decomposition is to reduce the number of variables and constraints within each small, isolated subnetwork to create a small optimization problem with minimum interaction requirements. After decomposition of the sensitivity matrix $A$, each subnetwork maintains the voltage levels in its subnetwork within certain limits by calculating the optimal adjustments to the involved DG outputs based on the information provided by the resulting decomposed matrix, as discussed below.

### B. Descriptions of Agents

The epsilon decomposition algorithm is implemented using an MAS architecture, thereby allowing the agents in a large network to group into small subnetworks with the communication links between agents being autonomously determined. To deploy this proposed approach, the system contains four types of agents: epsilon decomposition (ED) agent, violation detection (VD) agent, linear programming solver (LPS) agent, and distributed generation (DG) agent. Each agent uses its knowledge and behavior to autonomously manage its own activities and coordinate with only the appropriate agents while maintaining stable state of the system as described below.

#### 1) Epsilon Decomposition (ED) Agent:

The ED agent applies epsilon decomposition to the sensitivity matrix $A$ in (2) and activates LPS and VD agents, informing them about the other agents in the same subnetwork so that each agent links only with other agents within the same subnetwork.

As an example of decomposition, let us consider the sensitivity submatrix $A_{VQ}$ in (2), which can be expressed as

$$A_{VQ} = A'_{VQ} + \varepsilon \cdot B \qquad (4)$$

where $A'_{VQ}$ is the decomposed submatrix, which retains the strong couplings represented by elements greater than $\varepsilon$, and $\varepsilon \cdot B$ contains weak couplings. The resulting submatrix defines the topology of the subnetworks and the influence range of each DG. After decomposition, it is possible that not all DGs in a subnetwork will have strong couplings with all nodes in this subnetwork. Thus, the "range of influence" of a DG is defined as the nodes within its subnetwork for which the voltages can be affected by the output of the DG. This influence range can be contracted or expanded by adjusting the threshold value.

The ED agent selects and updates the $\varepsilon$ value based on its knowledge or when triggered by other agents or changes in the network. For example, this can be triggered by an LPS agent to request involving more DGs in the control when a DG has tripped, or when this DG is back online to restore the previous operation before the trip event. Moreover, this can be initiated by the network changes when a new DG unit is installed. This resets and updates the subnetworks and the ranges of influence of the DG in accordance with (2) and (4). The agents regroup and self-organize to adapt to the new network conditions without a complete re-engineering of the overall MAS framework. This process can occur at any time without stopping or restarting the MAS platform and its agents.

#### 2) Violation Detection (VD) Agent:

The VD agents monitor the status of busses and have knowledge of the voltage violation constraints, as specified by the voltage upper and lower normal operating limits (between 0.95 pu and 1.05 pu in this paper). When a VD agent observes a voltage violation at its bus, it requests the LPS agent in the same subnetwork to resolve the network voltage issue.

#### 3) Linear Programming Solver (LPS) Agent:

Each LPS agent acts as a control agent for its subnetwork. When an LPS agent receives a violation message from a VD agent in its subnetwork, it calculates the optimal adjustments to the generation of the involved DGs using a linear programming (LP) algorithm that is integrated within each LPS agent. Thus, the LPS agent has knowledge of: a) the DG agents in the same subnetwork and the surplus capacity of each DG; b) the sensitivity submatrix of the subnetwork that is used to determine how each DG affects the voltages of the nodes within its range of influence; c) the sensitivity coefficients, obtained from (2), that define how a DG influences the voltage magnitudes and phase angles of the primary and secondary sides of the network transformers; and d) the acceptable voltage limits (in this paper, 0.95–1.05 pu). The third constraint ensures that the network protectors used to avoid reverse active power flows through the network from the secondary side to the primary side are not tripped, as shown in (7) and (9). This approach can be applied to networks with DGs operating in the unity power factor (UPF) mode by increasing or decreasing the active power output and to networks with DGs operating in the power factor control (PFC) mode by injecting or absorbing reactive power.

It follows that when all DGs are operating in the PFC mode, the LPS agent can control the voltage optimally by minimally decreasing or increasing the reactive power outputs of the local DGs involved. For the LP problem in the PFC mode, the objective function is

$$\text{Max: } Min\{x_i\} \text{ (to control overvoltage)} \qquad (5)$$

$$\text{Min: } Max\{x_i\} \text{ (to control undervoltage)} \qquad (6)$$

subject to the following constraints:

$$\begin{cases} V_l \le V_0 + A_{VQ} \cdot x \le V_u \\ x \le Q_{Sur} \\ 0 \le \theta_{p0} + A_{\theta_p Q} \cdot x - (\theta_{s0} + \theta_{shift} + A_{\theta_s Q} \cdot x) \end{cases} \qquad (7)$$

When operating in the UPF mode, DGs can generate only active power. The LPS agent calculates the optimal generation adjustments for the involved DG to control the voltage with the following objective function:

$$\text{Max: } Min\{x_i\} \qquad (8)$$

s.t.

$$\begin{cases} V_l \le V_0 + A_{VP} \cdot x \le V_u \\ x \le P_{Sur} \\ 0 \le \theta_{p0} + A_{\theta_p P} \cdot x - (\theta_{s0} + \theta_{shift} + A_{\theta_s P} \cdot x) \end{cases} \qquad (9)$$

To solve this LP problem in the same way as the standard LP problem, we can add a slack variable $y$ to bring it into the same form as the standard LP problem. Let us consider overvoltage control in the PFC mode as an example:

$$\text{Max: } y \text{ (to control overvoltage)} \qquad (10)$$

s.t.
$$\begin{cases} V_l \leq V_0 + A_{VQ} \cdot x \leq V_u \\ x \leq Q_{Sur} \\ 0 \leq \theta_{p0} + A_{\theta_p Q} \cdot x - \left(\theta_{s0} + \theta_{shift} + A_{\theta_s Q} \cdot x\right) \\ x_i \geq y \ (i = 1\sim n, n \text{ DG agents are involved}) \end{cases} \quad (11)$$

In the above equations, $x_i$ is the power generation adjustment of the $i$-th DG agent; $x$ is the vector of all $x_i$; $V_u$ and $V_l$ are the upper and lower voltage limits, respectively; $Q_{Sur}$ and $P_{Sur}$ are the surplus capacities of the DG; $\theta_{p0}$ and $\theta_{s0}$ are the initial values of the network transformer voltage angles on the primary side and secondary side, respectively; $A_{\theta_p Q}$ and $A_{\theta_s Q}$ are the sensitivity submatrices of the reactive power adjustments and the transformer voltage angles on the primary and secondary sides, respectively; and $A_{\theta_p P}$ and $A_{\theta_s P}$ are the sensitivity submatrices of the active power adjustments and the transformer voltage angles on the primary and secondary sides, respectively.

Once an LPS agent receives a violation message from a VD agent, it first determines which of the DGs of its subnetwork are the "closest" to this violation, and then, considering only these involved DGs, it seeks a solution by employing the LP algorithm. The closest DGs to a node are defined as the neighboring DGs that can influence the node's voltage by adjusting their outputs, as determined by the range of influence of the DGs and as expressed in (4). Thus, when the LPS agent solves the LP problem in its subnetwork, only the closest DG agents to the violating node participate in voltage control, which further reduces the size of the optimization problem and the interaction requirements. The voltage control also becomes more "local" within each subnetwork, i.e., the VD agents coordinate with their local LPS agent to optimize only the local closest DG agents in the subnetwork.

After determining how to resolve the voltage violation problem, the LPS agent communicates the control adjustments to the involved DG agents to request an increase or decrease in the DG outputs and to restore the voltage of the subnetwork to within the normal operating limits. However, if the subnetwork is unable to resolve the voltage issue (e.g., a DG is not available or has lost communication), it will request the ED agent to update the $\varepsilon$ value to involve more DG agents in the problem. As a result, the control agents reconfigure and self-organize in the new identified subnetworks to regulate the voltage.

*4) Distributed Generation (DG) Agent:*

Each DG agent represents its DG that is connected to the distribution network and performs control actions to adjust its output. DG agents receive generation adjustment signals from the LPS agent in the same subnetwork to maintain appropriate voltage levels.

The DG agent has knowledge of its constraints (such as generation capacity) and the decomposed sensitivity $A'$ matrix in order to dynamically update its range of influence. It also shares its constraints and availability with other agents. For instance, if a DG is not available (e.g., disconnected from the network), the DG agent informs the LPS agent so that the LPS agent can find solutions using other DGs through the self-organizing mechanism. When the DG is back online, it informs the LPS agent sharing its availability and constraints.

It is also noted that the DG agent shares its constraints, such its available surplus capacity, with the LPS agent in order for the LPS agent to consider when solving the LP problem. This is used by the LPS agent to maintain the operation of the involved DGs and the voltage regulation process, and to enable the system to adapt based on conditions and constraints of involved DGs and network events as summarized next.

*C. Implementation for Distributed Voltage Regulation*

The ED agent decomposes the sensitivity matrix with the initial $\varepsilon$ value as shown in (4). The number of subnetworks depends on the $\varepsilon$ value, which is selected and updated autonomously in accordance with the purpose of decomposition, i.e., the desire to achieve smaller subnetworks for distributed control. The $\varepsilon$ value is initialized with the value that yields the largest number of groups.

The LPS and VD agents receive the required knowledge of the decomposed network, including which agents are in the same subnetwork, so that the agents only need to communicate with other agents in the same subnetwork. After decomposition, the agents within each subnetwork coordinate to realize and maintain distributed voltage regulation. If a subnetwork cannot regulate the voltage using the involved DGs (e.g., a DG has tripped or does not have enough surplus capacity), the agents will coordinate and self-organize to involve more DG agents in the control problem.

The process and steps of implementing the algorithm are summarized below.

*Step 1):* The ED agent initiates the system via (4) and activates the LPS and VD agent according to the subnetworks identified.

*Step 2):* Each VD agent starts monitoring its bus for voltage violations; if such a violation occurs, it sends a violation message to the LPS agent of its subnetwork.

*Step 3):* The LPS agent checks the violation message received from its subnetwork and identifies generation adjustments for the involved DG agents, as expressed in (5)-(7) for the PFC mode or in (8) and (9) for the UPF mode.

*Step 4):* The LPS agent then sends the control actions to the DG agents to adjust their output.

*Step 5):* After the voltage is normalized in the subnetwork, the agents return to Step 2 to continue monitoring and controlling the system.

*Step 6):* If the subnetwork cannot regulate the voltage (e.g., an involved DG has tripped), the LPS agent will request the ED agent to determine a new $\varepsilon$ value for decomposition and involve more DG agents in the control problem. Subsequently, the agents will self-organize to regulate the voltage.

*Step 7):* When the network returns to normal operation (e.g. the DG is back online), the agents reorganize and return to Step 2 to continue monitoring and controlling the system.

In addition to the above steps, the system is able to adapt to network changes, such as expansion of the network or the removal of DGs through its self-organizing mechanism. The system will dynamically update the decomposition of the network, and the agents organize into new subnetworks.

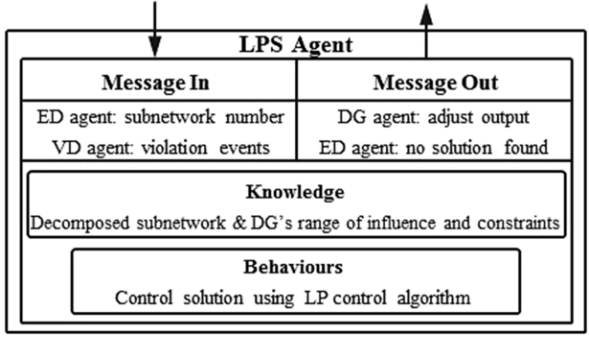

Fig. 1. LPS agent architecture in Presage2.

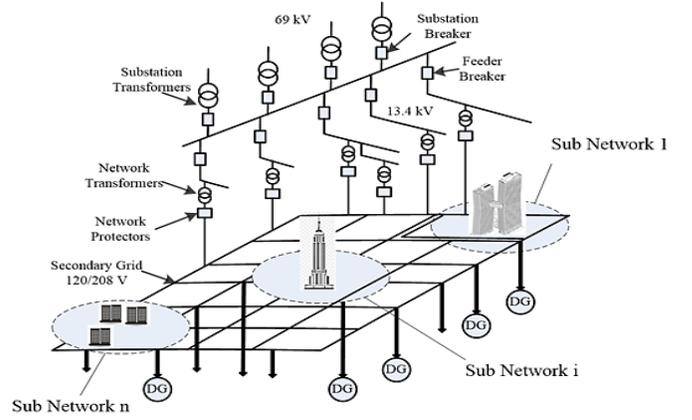

Fig.2. Test system demonstrating subnetworks after decomposition.

## IV. APPLICATION TO A HEAVILY MESHED DISTRIBUTION NETWORK

### A. Self-Organizing MAS Implementation

The MAS architecture was deployed using the Simulation of Agent Societies 2 (Presage2) framework [24], which offers agent communication capabilities and improved autonomy. Presage2 provides the flexibility to design self-organizing MASs able to meet the requirements of electrical network control and management applications. An example of selected knowledge and behavior of an LPS agent within Presage2 and its interactions with other agents is illustrated in Fig. 1.

The message format in Presage2 is as follows.

*message (performative, sender, destination, time, content)*

The message *performative* identifies the purpose of the message, for instance "*inform*", "*confirm*", and "*query-if*" as described in the FIPA-ACL standard [25]. The message *sender* and the message *destination* denote the agents that are sending and receiving the message, respectively. The message *time* is the time at which the message was sent, while the message *content* carries information that is sent between the agents. A message example is as follows.

*message (inform, VD$^{362}$, LPS$^{34}$, 23, V<0.912>)*

An increasing number of smart grid and power systems applications require more frequent and improved data communications. There are extensive developments and research taking place in this field for smart grids, and the future communications systems will underpin the advanced functionality of this paper. The communication networks under the smart distribution systems paradigm provides connectivity to many devices distributed throughout a particular geographical region. A detailed description of these information and communication technologies (ICTs) and infrastructure to support the operation of smart distribution networks can be found in [26]-[28]. Moreover, future smart grids propose the use of advancements in ICT such as IoT [29] and energy internet [30], and with the support of future ICT technologies such as 5G [31]. However, a detailed examination of communications requirements and their implementation are beyond the scope of this paper.

It is worth noting that we assume instantaneous communication between agents when a voltage regulation event occurs. However, in such an event-based system, it is important to request control and communication resources instantaneously. To maintain the availability of resources, prediction protocols based on network conditions [32] can be implemented to improve the proposed method. Developing such predictive rules (e.g., due to a disturbance) for agents and their use of communication resources was investigated recently in [32]. In our system, rules can be implemented based on network or other agents events and conditions, such as a DG trip or available surplus capacity. This can enable the processing or communication system to make unused resources available to other agents or to reconfigure the sleeping mode to save energy. Future research will be carried out in this direction.

### B. The Test System

The test system is a model of a real heavily-meshed secondary network containing 2083 nodes, 224 network transformers (13.8 kV to 480 V or 216 V), and 311 PQ loads. As shown in Fig. 2, the primary feeders are at 13.8 kV and contain 1043 nodes, while the secondary network contains the remaining 1040 nodes at 216 V or 480 V. In this network, the 311 DGs are installed at the load buses in the secondary network. The largest network in the secondary grid has all of its transformers connected on the secondary side, which creates a heavily-meshed network of 284 loads. The decomposition of the network is demonstrated in Fig. 2.

Table I shows how the value of ε influences the number of control subnetworks. For example, a smaller value of 0.004 results in fewer groups than a larger value of 0.016, while a larger ε value provides more small-size subnetworks, which results in a smaller control problem. However, a large ε value may also yield suboptimal control strategies as a result of neglecting links with some significance. The MAS architecture is deployed in the model system, and the agents self-organize into subnetworks defined by the communication links between agents.

TABLE I
RESULTS OF EPSILON DECOMPOSITION WITH VARIOUS ε VALUES FOR DGS OPERATING IN PFC AND UPF MODES

| ε | DG with PFC mode | DG with UFC mode |
|---|---|---|
| | No. of DG subnetworks | No. of DG subnetworks |
| 0.004 | 19 | 20 |
| 0.006 | 24 | 35 |
| 0.008 | 34 | 47 |
| 0.010 | 57 | 65 |
| 0.012 | 82 | 75 |
| 0.014 | 80 | 72 |

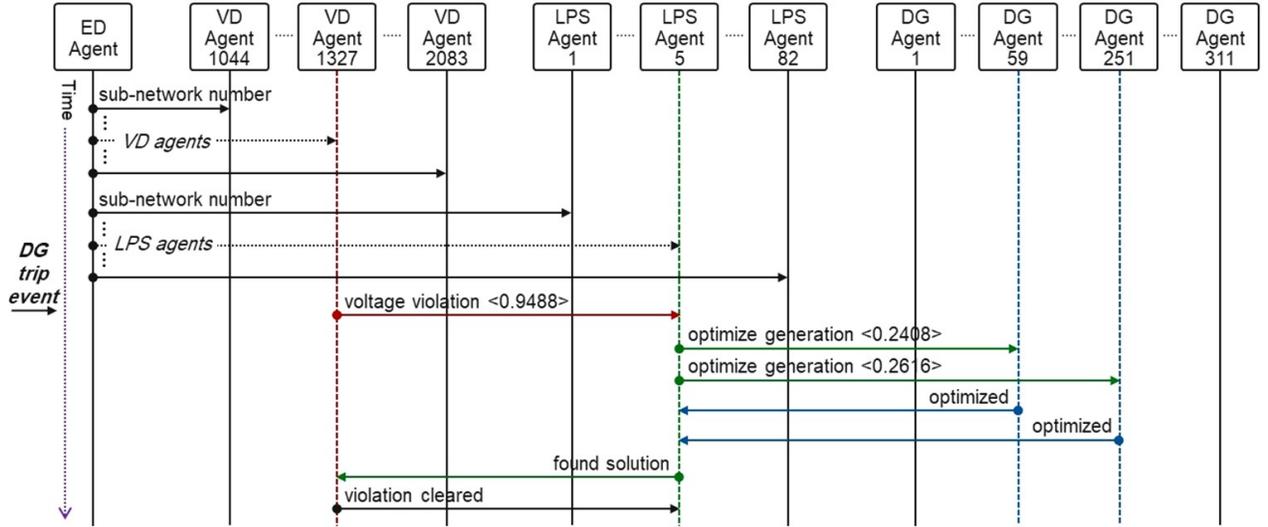

Fig. 3. Communication among agents for distributed voltage regulation (Case Study 1).

TABLE II
RESULTS OF CONTROL WITH $\varepsilon = 0.012$ IN PFC AND UPF MODES FOR THE SAME VIOLATION EVENT (CASE STUDY 1)

| Control Mode | PFC | UPF |
|---|---|---|
| No. of DG agents in the subnetwork | 81 | 25 |
| No. of DG agents involved in the control | 2 | 1 |
| No. of involved nodes in the control | 21 | 8 |
| $P_{loss}$ after control (pu) | 0.2507 | 0.2330 |
| $Q_{loss}$ after control (pu) | 0.8332 | 0.8061 |

It is noted that when the network is going under high penetration of distributed renewable resources, intermittent renewables may increase the voltage violations events in the networks. However, to reduce the pressure on the overall network and maintain the performance of the control method, we assume that DGs have a mechanism (such as using energy storage [33]) to help reduce variations of generation. This can also offer dispatchable resources available to use by the proposed distributed voltage regulation scheme.

## V. RESULTS AND DISCUSSION

In the following, we present the simulation results of the proposed distributed control algorithm performed in the test system under various network conditions to validate the autonomy and adaptability of the system. In addition, to verify the overall performance of the proposed framework, we compare the results with four other control approaches: i) no DG control, ii) using our self-organizing method without partitioning, iii) the local voltage control only without coordination where each DG acts on its local bus voltage only [34], [35], and, iv) the community detection algorithm based on the modularity index method presented in [36]. It is worth noting that when implementing our presented method without partitioning, it uses the voltage sensitivity coefficients associated with the whole network; therefore, it can be used as a centralized equivalent control case benchmark since its LP problem accounts for all the links between nodes and DGs in the secondary network. For the community detection algorithm, the partitioning method is based on the sensitivity matrix (5) and the voltage is regulated using DGs in each partition.

### A. Case Study 1: Distributed Voltage Regulation

The algorithm is simulated in the network to demonstrate effective distributed voltage regulations with $\varepsilon = 0.012$ in PFC and UPF modes. For PFC mode, the agents in the secondary network are decomposed into 82 isolated subnetworks, with one LPS agent activated in each subnetwork. As demonstrated in the agent interaction chart in Fig. 3, the ED agent informs the VD and LPS agents about their subnetworks. To create a voltage violation, we assume a disconnection of DG unit 71, in subnetwork 5, from the secondary network, which causes a voltage violation to appear in the distribution network. The violation is detected by VD agent 1327, in subnetwork 5, with a voltage value of 0.9488 pu. This VD agent sends a message with the location and value of the voltage to the corresponding subnetwork's control agent, LPS agent 5, as illustrated in Fig. 3.

For this case study, subnetwork 5 is the largest DG group, containing 81 DGs and around 250 nodes of the total 311 DGs and 1040 nodes in the secondary network. Although the size of this subnetwork is much smaller than the original secondary network, solving the LP problem may still have 81 variables and around 500 constraints if all the DGs of this subnetwork are involved in the voltage regulation function. However, for such a large subnetwork, it is possible that not all DGs have a strong coupling with all nodes in the subnetwork. Thus, after decomposition, only the corresponding closest DG or DGs whose range of influence covers the node voltage are involved in the voltage regulation process. Therefore, LPS agent 5 first determines the DG agents in its subnetwork that can influence the node voltage, which are DG agent 59 and DG agent 251 in this case, as shown in Fig. 3. As a result, out of the 81 DG agents in this subnetwork, only two DGs are involved in the control problem. This further reduces the size of the LP control problem from 81 variables and nearly 500 constraints, to only 2 variables and 42 constraints, as show in Table II. It also simultaneously reduces the interactions required between only the involved agents. As shown in Fig. 3, when LPS agent 5 finds the solution, it sends messages to DG agents 59 and 251,

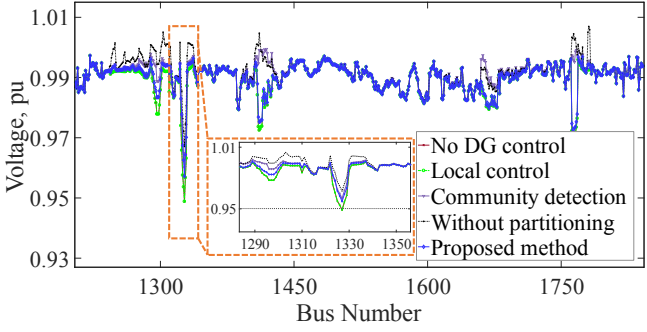

Fig. 4. Voltage profiles of the secondary network under different scenarios.

TABLE III
COMPARISON OF THE CONTROL PARAMETERS AND RESULTS

| Technique | Proposed Method | Without Partitioning | Local Control [34] | Community Detection [36] |
|---|---|---|---|---|
| No. of involved DGs | 2 | 12 | 1 | 9 |
| No. of involved nodes | 21 | 138 | 1 | 123 |
| $P_{loss}$ after control (pu) | 0.2507 | 0.2667 | 0.2512 | 0.2594 |
| $Q_{loss}$ after control (pu) | 0.8332 | 0.8395 | 0.8360 | 0.83735 |

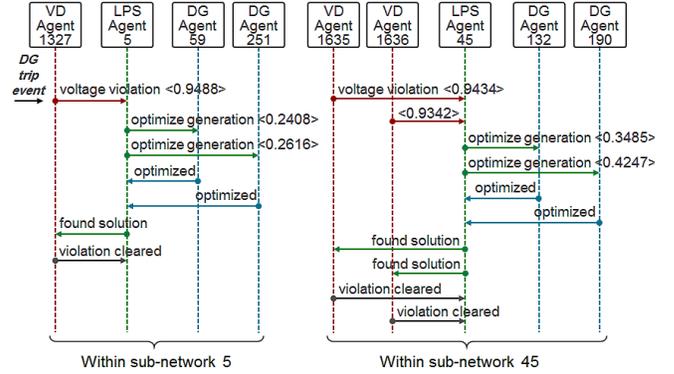

Fig. 5. Coordination among agents to simultaneously and independently control the voltage within each subnetwork (Case Study 2).

instructing them to inject reactive powers of 0.241 pu and 0.262 pu, respectively.

To test the algorithm in UPF control mode, the same voltage drop as above is analyzed. Table II compares the results of voltage regulation under the different control modes for the same $\varepsilon$ value. It also shows that in the UPF mode, fewer DGs and nodes are involved in solving the control problem, with lower system power losses.

This case study demonstrates effective distributed voltage regulation based on the proposed MAS architecture for a particular area with sufficient local DG. This architecture may also have further applications for static partitioning when considering proper DG allocation planning in distribution networks [8], [37], which can be used as a reference to expand the network.

### B. Performance Comparison

We use case study 1 to compare the performance of presented distributed voltage regulation to that of the system without DG control, the basic control on local bus voltage only without coordination, the non-partition control scheme, and the community detection algorithm. In this case study, we compared the results when using the community detection algorithm to evaluate the performance of our proposed distributed voltage regulation and when using only the closest DG.

Fig. 4 illustrates the voltage profiles of the nodes after the control actions for all compared methodologies. The horizontal dotted line in the inset in Fig. 4 represents the lower voltage limit (0.95 pu), which is exceeded at bus 1327. The best solution could be achieved by the global optimization method without partitioning the network (black dotted line), which regulates the voltage by solving the control problem considering all links between DGs and nodes in the network.

Fig. 4 shows that the proposed method (blue line) is as effective as the method without partitioning, while having less influence on other nodes when compared to no partition and to the community detection algorithm. When the community detection algorithm (purple line) is applied, the effected partition uses all DGs in its partition when solving the control problem and regulating the voltage, while our proposed methodology seeks to use the closest DGs only, as summarized in Table III. This table indicates that when compared to no partition and the community detection algorithm, in our methodology fewer DGs are involved in control of the voltage, and the size of the control problem is reduced significantly. When the uncoordinated local control method (green dotted line) is applied, the voltage at bus 1327 is below the 0.95 pu limit. This is because it does not have a coordination mechanism to regulate the voltage when the corresponding local DG 71 at bus 1327 has tripped.

The final solution of the proposed method may not lead to the same result obtained by the overall optimization of the entire secondary network without partitioning, but, as shown in Fig 4, they are comparable. However, the aim of the proposed technique is to provide a voltage regulation platform that uses distributed approach with sufficiently accurate results, while dealing with increasing complexity of such large networks with DGs and a need to obtain fast solutions In Table III, we also report the power losses in the secondary network after control. This can be seen as another criterion showing that the proposed method maintains power losses comparable to solving for the whole network, and also may result in reduced power losses such as in this case study.

### C. Case Study 2: Voltage Violations in Multiple Subnetworks

The agent communication diagram in Fig. 5 shows the agents involved in this control problem. To study the control algorithm, it is assumed that there is no communication delay, such that each LPS agent receives the violation messages at the same time; i.e., each of LPS agent 5 and LPS agent 45 receives violation messages simultaneously from the VD agents in its subnetworks. Each LPS agent identifies solutions for its own subnetwork to optimize the involved DG outputs and to control the voltage in its area. Therefore, the voltage control problem is divided into 2 LP problems that are solved independently by each corresponding LPS agent. The information of these two LPS agents to solve their LP problems is summarized in Table IV.

| Area | Subnetwork 5 | Subnetwork 45 |
|---|---|---|
| LPS agent ID | 5 | 45 |
| No. of DG agents in the subnetwork | 81 | 3 |
| No. of DG agents involved in the control | 2 | 2 |
| No. of nodes involved in the control | 21 | 4 |

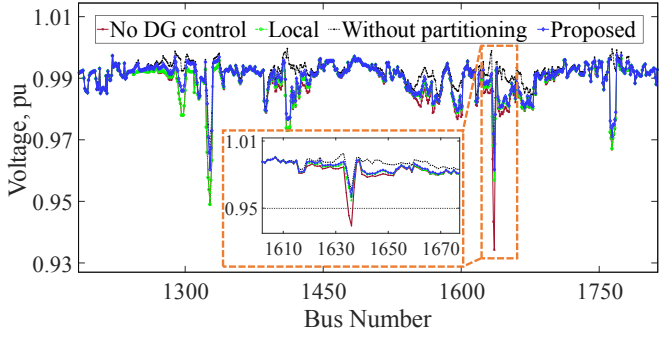

Fig. 6. Voltage profiles of busses showing violations and regulation in multiple subnetworks at the same time (Case Study 2).

Simulations were performed to test the performance of the system when multiple voltage violation events occur in different subnetworks at the same time. This is simulated by simultaneously tripping multiple DGs in different subnetworks, causing voltages to exceed the acceptable operating limits in each affected subnetwork. In this case, in addition to the disconnection of DG 71, as in the first case study, we assume that a disconnection of a second DG (DG 119 in subnetwork 45) occurs at the same time. As a result, as shown in Fig. 5, in addition to a voltage drop appearing in subnetwork 5 and being detected by VD agent 1327 (similar to Case Study 1), violation events are detected by VD agents 1635 and 1636 (in subnetwork 45), which communicate these violations to LPS agent 45.

The voltage profiles of the secondary network after control for various methods are presented in Fig. 6. The voltage at the effected buses 1635 and 1636 in subnetwork 45 before and after control is shown in Fig. 7. The performance is compared with the method without partitioning and when using basic local control. As expected, the results show that the voltage regulation of the proposed distributed method is almost equivalent to the method without partitioning. As shown in Fig. 6, the voltage limits in the proposed method are maintained in each subnetwork with very little influence on other areas. This is because the proposed distributed voltage regulation enables each subnetwork to regulate the voltage independently and to use the closest DGs to a node.

As summarized in Table V, in comparison to the situation where the network is not partitioned, the proposed method enables solving the control problem with less DG involved and with a reduced size of the control problem. This not only leads to results comparable to solving for the whole network, but also allows dividing and simplifying the problem into smaller independent sub-problems. In this case study, the basic local control without coordination was able to successfully regulate the voltage in subnetwork 45 using DG number 190 connected to bus 1635.

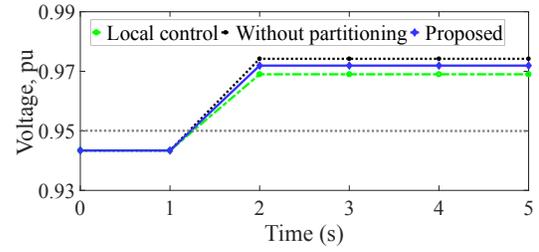

(a)

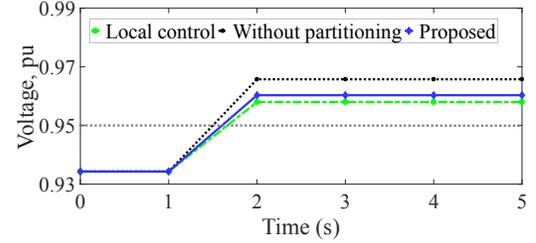

(b)

Fig. 7. Voltage of nodes in subnetowrk 45. (a) Bus 1635 . (b) Bus 1636.

| Technique | Proposed Method | Local Control | Without Partitioning |
|---|---|---|---|
| No. of involved DGs | 4 | 3 | 15 |
| No. of involved nodes | 25 | 3 | 271 |
| $P_{loss}$ after control (pu) | 0.3263 | 0.3255 | 0.3402 |
| $Q_{loss}$ after control (pu) | 0.9076 | 0.9105 | 0.9161 |

### D. Case Study 3: The Self-Organizing Mechanism

The self-organizing property of the system is implemented by autonomously adjusting the decomposition of the subnetworks to adapt to anomalies. For $\varepsilon = 0.012$, with the DGs operating in the UPF mode, 75 subnetworks are generated. A disconnection of DG 192 in subnetwork 51 is considered in this test. As illustrated in the communication diagram in Fig. 8, VD agent 1694 (in subnetwork 51) reports a voltage violation event to LPS agent 51, which identifies two DG agents in its network, namely, DG agents 192 and 247, that can influence the voltage of the violating bus. However, because DG agent 192 was disconnected from the network, the subnetwork attempts to control the voltage with the one available DG agent (247).

In this scenario, the subnetwork is not able to restore the voltage using the available DG resources of its subnetwork and informs the ED agent to select a smaller $\varepsilon$ value to expand the ranges of influence of the DGs and to involve more DG agents in the voltage control problem. When selecting $\varepsilon=0.01$, the network is reset, and the agents regroup and reorganize based on the new decomposition. As shown in Fig. 8, VD agent 1694 is now in subnetwork 9, in which four DGs can influence the bus voltage. Thus, the corresponding control agent, LPS agent 9, finds a solution and sends the optimal adjustment messages to the DG agents to take control actions. Due to the limited space available for the interaction diagram,

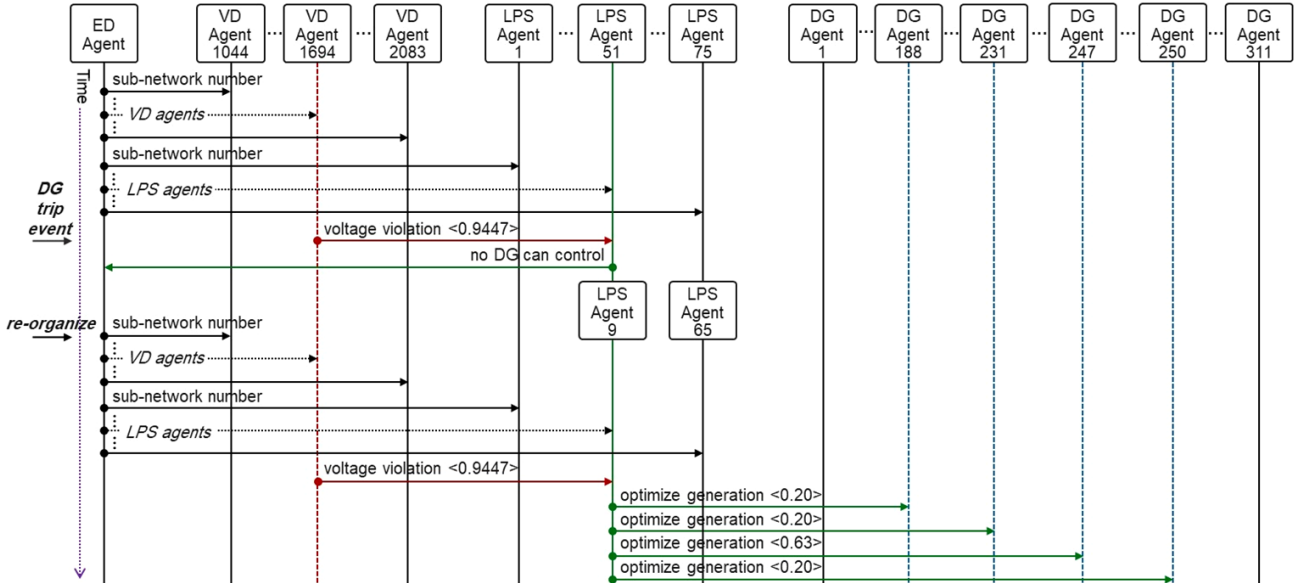

Fig. 8. Coordination and self-organization of agents for distributed voltage regulation (Case Study 3).

TABLE VI
RESULTS OF THE VIOLATION EVENT WITH THE SELF-ORGANIZING
MECHANISM (CASE STUDY 3)

| Control Mode | UPF | |
|---|---|---|
| ε value | 0.012 | 0.010 |
| No. of DG agents can influence the node | 1 | 4 |
| Sufficient DG resources | No | Yes |

TABLE VII
COMPARISON OF THE CONTROL PARAMETERS FOR CASE STUDY 3.

| Technique | Proposed Method | Local Control | Without Partitioning |
|---|---|---|---|
| No. of involved DGs | 4 | 1 | 7 |
| No. of involved nodes | 25 | 1 | 146 |
| $P_{loss}$ after control (pu) | 0.2640 | 0.2840 | 0.2722 |
| $Q_{loss}$ after control (pu) | 0.8427 | 0.8658 | 0.8495 |

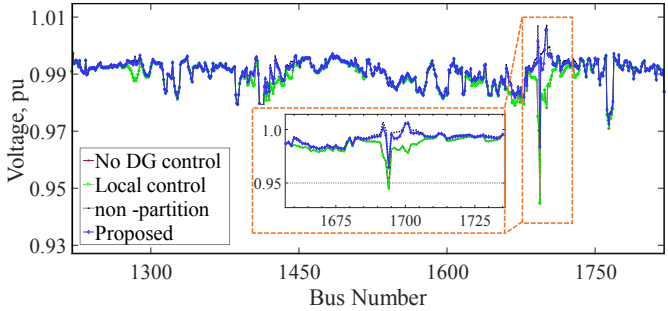

Fig. 9. The voltage profiles after regulation through self-organization.

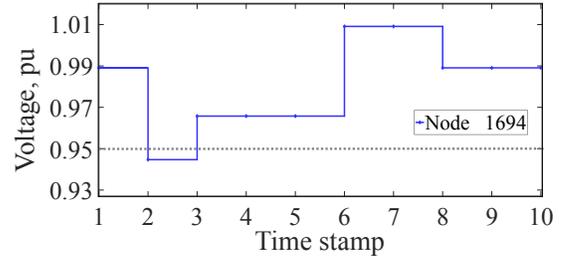

Fig. 10. The voltage levels at bus 1694 realized through self-organization.

Fig. 8 shows only the agents that are involved in this control problem. Table VI also summarizes the proposed distributed control results for scenarios with different ε values for the same violation event.

As presented in Fig. 9, after the self-organizing mechanism and the outputs are optimized, the voltage is controlled and returned to an acceptable level. The performance of various methods is also included in Fig. 9 and Table VII. Figure 9 shows that the proposed mechanism can perform as effectively as the system without partitioning. Although the basic local control does not require any communication, it does not have a coordination mechanism to adapt and regulate the voltage when the involved local DG has tripped. Table VII summarizes the results of the methods under study. They are consistent with the conclusions of case studies 1 and 2.

The system is not only capable of maintaining a stable state, e.g., by expanding the network when DG resources are not available, but can also self-divide into a more distributed state of small groups once the network returns to normal operation (e.g. when a DG is back online in this cases study). As illus-trated in Fig. 10, in this case study, DG 192 trips at time T2 causing undervoltage at bus 1694. The agents self-organize at time T3 to regulate the voltage successfully. When DG agent 192 returns back online at T6, it informs the LPS agent in order to return to the previous operation before the DG trip. The LPS agent first resets the four DGs agent (used to regulate the voltage due to the trip of DG 192) at T8 to return to normal operation. The LPS agent then informs the ED agent to restore the initial ε value before the DG trip event. The system self-reorganizes, and the ε value is reset to its value before the DG tripped. This mechanism also allows the system to self-organize in response to various conditions such as uncertainty and the availability of energy resources over time (e.g., DG does not have sufficient power or has lost communication).

## VI. CONCLUSIONS

This paper presents a self-organizing distributed voltage regulation approach that can be applied in large networks and uses only the appropriate neighboring DG. The distributed control technique is implemented using an MAS framework, in which the agents autonomously group themselves into sub-networks to control the voltage through local interactions of agents in a cooperative way. Additionally, the system can realize the adjustment of DG active power outputs in UPF mode or reactive power outputs in PFC mode. The system can also adapt to time-varying network conditions to maintain stable control by dynamically updating the grouping of the network without re-engineering the complete system.

The effectiveness of the proposed approach is validated through simulations based on a complex network model of a real heavily-meshed secondary network, and the performance is compared to various techniques. Simulation results prove the autonomy of subnetworks to control the voltage independently using only the involved DGs. The system was also tested to validate the adaptability and robustness of the system by maintaining stable voltage control in response to network anomalies over time.

This method offers a novel MAS-based distributed voltage regulation approach with improved self-organization capabilities for complex distribution networks, potentially giving rise to an adaptive and robust control approach suitable for wider adoption in smart grids. The future work includes extending the presented method to explore and solve voltage control challenges, such as voltage collapse analysis, posed by the increasing penetration of distributed and renewable generation sources in smart grids.